\documentclass[aps,reprint,prd,nofootinbib]{revtex4-1}

\usepackage{bm}
\usepackage{graphicx} %
\usepackage{amsmath} %
\usepackage{amssymb} %
\usepackage{url}
\usepackage{subfigure} %
\usepackage{float} %
\usepackage{upgreek} %
\usepackage{multirow} %
\usepackage{color} %
\usepackage{lineno} %
\usepackage{hyperref} %
\usepackage{booktabs}
\usepackage{placeins}
\usepackage{ulem}
\usepackage[usenames,dvipsnames]{xcolor}
\hypersetup{colorlinks=true, urlcolor=black, linkcolor=blue, citecolor=red} %
\usepackage{amsmath}
\usepackage{tcolorbox}
\allowdisplaybreaks[4]

\begin{document}

\title{Cosmic microwave background spectral distortions \\
from Rayleigh scattering at second order}

\author{Atsuhisa Ota${}^{1}$}
\email{iasota@ust.hk}

\affiliation {${}^1$HKUST Jockey club Institute for Advanced Study, The Hong Kong University of Science and Technology, Clearwater Bay, Hong Kong, P.R.China } 

\date{\today}

\begin{abstract}
Cosmic microwave background~(CMB) spectral distortion from Rayleigh scattering is calculated for the first time in rigorous second-order cosmological perturbation theory.
The new spectral distortion is sensitive to acoustic dissipation at $10^{-2}<k{\rm Mpc}/h<1$, which slightly extends the scale constrained by the CMB anisotropies.  
The spectral shape is different from either temperature perturbations or any other traditional spectral distortions from Compton scattering, such as $y$ and $\mu$.
The new spectral distortion is not formed in the late Universe, unlike the thermal Sunyaev-Zel'dovich effect degenerated with the primordial $y$ distortions since photons must be hot for Rayleigh scattering.
Therefore, ideal measurements can distinguish the signal from the other effects and extract new information during recombination.
Assuming cosmological parameters consistent with the recent CMB anisotropy measurements, we find the new spectral distortion is $6.5\times 10^{-3}$Jy/str, which is one order of magnitude smaller than the currently proposed target sensitivity range of voyage 2050.

%\pacs{04.80.Cc, 95.30.Sf, 98.70.Vc, 98.80.Es}

\end{abstract}

\maketitle

\section{Introduction}

%Successful measurements of the cosmic microwave background~(CMB) linear anisotropies have provided detailed information of cosmological parameters in the decades~\cite{Smoot:1992td,WMAP:2003ivt,Akrami:2018odb,Planck:2019evm,Planck:2019kim}.
%However, the linear anisotropies are theoretically and observationally exhausted except for the final effort for the polarization B-mode from primordial gravitational waves~\cite{Kogut:2011xw,PRISM:2013fvg,Ade:2018sbj,Kogut:2019vqh,Aravena:2019tye,Abazajian:2019eic,NASAPICO:2019thw,LiteBIRD:2020khw}. 
%Therefore, we need to go beyond the linear theory and linear observables to understand the Universe further.
Spectral distortions of the cosmic microwave background (CMB) are potential cosmological observables for the next decades~\cite{Zeldovich:1969ff, Sunyaev:1970eu, Hu:1992dc}.
Although cosmologists have prioritized the anisotropy measurements to spectrum measurements since COBE FIRAS reported the almost ideal black body spectrum in 1996~\cite{Fixsen:1996nj}, several observational projects are envisioned or ongoing targeting tiny spectral distortions~\cite{Kogut:2011xw,Kogut:2019vqh,PRISM:2013fvg}, including extreme measurements proposed for voyage 2050~\cite{Chluba:2019kpb,Chluba:2019nxa}.
The spectral distortions are formed due to various non-equilibrium phenomena in the Universe.
For example, acoustic dissipation of short-scale cosmological perturbations~\cite{Hu:1994bz,Chluba:2012gq,Chluba:2012we,Dent:2012ne,Ota:2014hha,Ota:2018zwm}, atomic processes during recombination~\cite{Wong:2005yr,Sunyaev:2009qn,Chluba:2015gta,Chluba:2008aw}, and 
energy injection from unknown sectors~\cite{Chluba:2009uv,Chluba:2011hw,Chluba:2013pya,Slatyer:2015kla,Ali-Haimoud:2021lka,Kumar:2018rlf,Acharya:2019owx,Choi:2017kzp} can introduce the distortions.
There is no universal prescription for non-equilibrium physics, so considering all possible spectral distortions and their cosmological implications in depth must be indispensable.

Photons interact with free electrons via Thomson scattering during recombination.
While we normally ignore electrons bounded in atoms, those are also scattered, which is known as Rayleigh scattering.
%Rayleigh scattering is also elastic as Thomson scattering, but the cross-section is frequency-dependent.
Rayleigh scattering is also elastic as Thomson scattering, but the cross-section is frequency-dependent.
%The bounded electrons are more likely to interact with photons whose energy is close to the energy gaps $h\nu_i$ of the electron states.
%For photons $\nu/\nu_i\ll1$ with the electron states $h\nu_i$, the Rayleigh cross section is Taylor-expanded and the leading term is $(\nu/\nu_i)^4$.
%$\nu/\nu_i\ll1$ is a valid approximation for photons during recombination, and including Rayleigh scattering can be important.
This is because the bounded electrons are more likely to interact with photons whose energy is close to the energy gaps $h\nu_i$ of the electron states.
The Rayleigh cross section is Taylor-expanded and behaves as $(\nu/\nu_i)^4$ during recombination. %, and including Rayleigh scattering can be important.

Ref.~\cite{Yu:2001gw} showed that Rayleigh scattering of neutral hydrogen induces additional scattering, which changes the damping scale in the CMB anisotropies.
Ref.~\cite{Lewis:2013yia} included neutral helium and polarization, and they presented a precise numerical calculation of the CMB angular power spectrum.
Then, Ref.~\cite{Alipour:2014dza} developed a moment expansion for the frequency dependence and estimated the back reaction of Rayleigh scattering to baryon acoustic oscillation.
Recently, the Rayleigh anisotropies are also considered for constraining primordial non-Gaussianity and neutrino mass in Refs.~\cite{Beringue:2020wxw, Coulton:2020oxw}.
Thus, previous works on Rayleigh scattering are about the linear anisotropies.

While the photon spectrum is considered to be a local Planck in the early Universe, a spectral distortion is secondarily sourced by Silk damping of small-scale anisotropies~
\cite{Pitrou:2009bc,Chluba:2012gq,Renaux-Petel:2013zwa,Pitrou:2014ota,Ota:2016esq,Pitrou:2019hqg}.
Physically speaking, acoustic oscillation of the viscous fluid leads to friction heat at second order, and the energy release distorts the fiducial Planck spectrum when the Universe is out of chemical equilibrium.
Including Rayleigh scattering, the acoustic source can also be frequency-dependent, and then Rayleigh scattering at second order will introduce a different spectral distortion.
This paper computes the spectral distortion generated by second-order Rayleigh scattering for the first time.
Based on a rigorous framework of the general relativistic Boltzmann equation, we compute the spectral distortion due to Rayleigh scattering and discuss the size, detectability, and ability to constrain cosmological models. 

We organize the paper as follows.
In section \ref{sec2}, we provide the Rayleigh scattering cross-section for neutral hydrogen, helium, and singly ionized helium numerically.
Then we review the first-order Rayleigh anisotropies in section \ref{sec3}.
Section \ref{sec4} is the main part of the paper. 
We derive the spectral distortion from Rayleigh scattering at second order.
We discuss several possible cosmological constraints on the new spectral distortions in Section \ref{sec5}.
Finally, we conclude the paper in Section \ref{sec6}.

\section{Scattering cross section}\label{sec2}

\begin{table}
\caption{Expansion factors for Rayleigh scattering.
The neutral hydrogen and the neutral helium parameters are taken from Ref.~\cite{Alipour:2014dza}, and singly ionized helium is obtained as $c^{(\rm HeII)}_{2i+4} =c^{(\rm H)}_{2i+4}/2^{8+4i}$ as discussed in Ref.~\cite{Fi_k_2016}.}

\label{table2}
\begin{tabular}{l ccc}
$2i+4$ & $c^{(\rm HI)}_{2i+4}$ & $c^{(\rm HeI)}_{2i+4}$ & $c^{(\rm HeII)}_{2i+4}$ \\
\hline  
4  & 1.265625 & 0.120798 & 0.00494385\\
6  & 3.738281 & 0.067243 & 0.000912666 \\
8  & 8.813931 & 0.031585 & 0.00013449 \\
10 & 19.153795 & 0.014153 & 0.0000182665\\
12 & 39.923032 & 0.006226  & 2.3796$\times 10^{-6}$
\end{tabular}	
\end{table}

\begin{figure}
\centering
  \includegraphics[width=\linewidth]{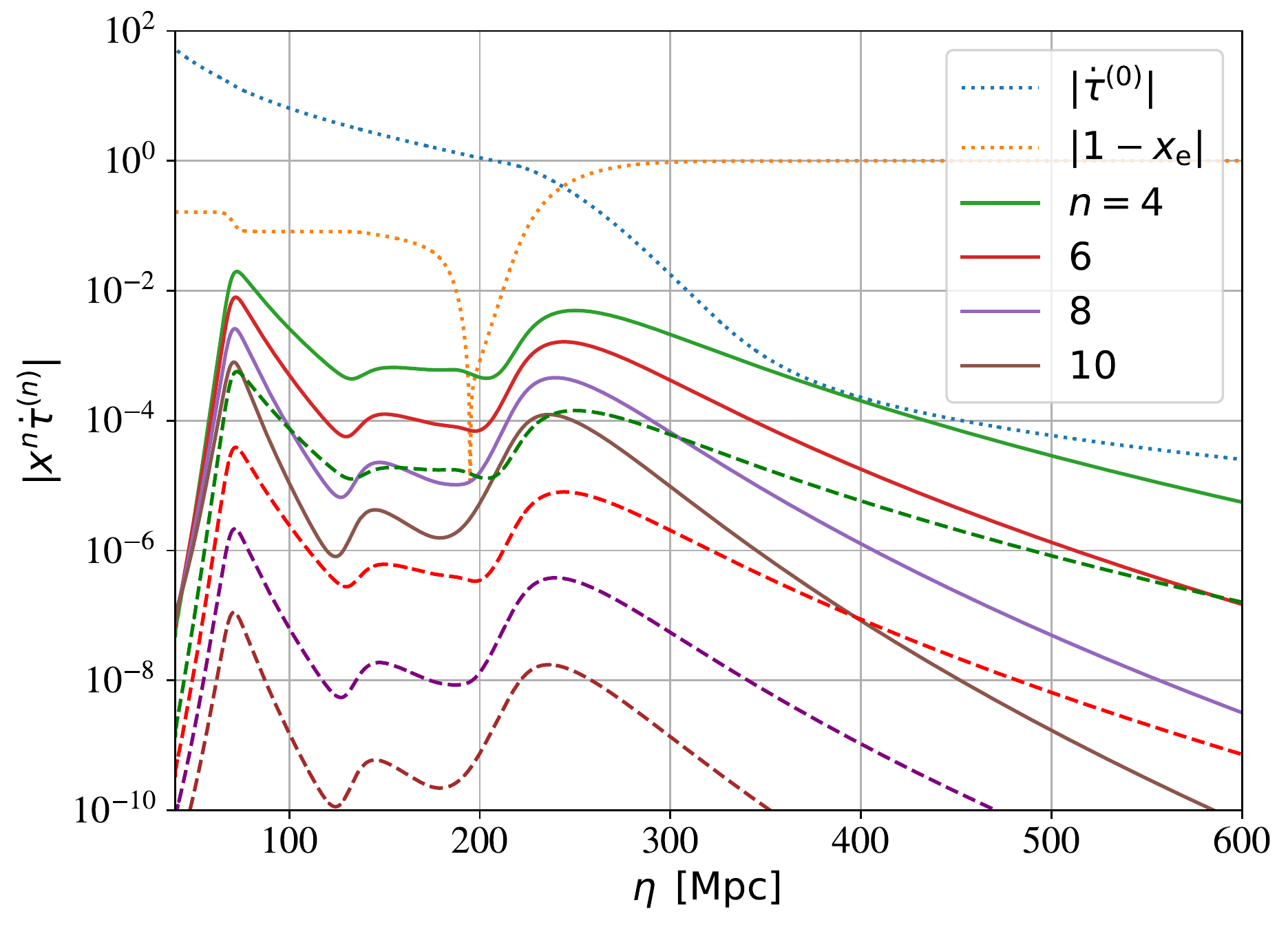}
  \caption{The differential optical depth multiplied by $x^n$ for various frequencies.
  The solid and dashed curves imply 857GHz and 357GHz, respectively.
  The peak at $\eta \lesssim 100 {\rm Mpc}$ is Rayleigh scattering of the singly ionized helium, which is bigger than the peak of neutral hydrogen. 
  }
  \label{recfast_out}
\end{figure}

During recombination, singly ionized helium~(HeII) is first produced and then neutral helium~(HeI) is formed.
In the last stage of recombination, the remaining electrons are captured in neutral hydrogen~(HI) atoms.
Rayleigh scattering starts after singly ionized helium appears at $z\sim 6000$.
The non-relativistic Rayleigh scattering cross section including those three species is expanded into~\cite{Alipour:2014dza}
\begin{align}
	\sigma_{\rm R} = \sigma_{\rm T}\sum_{j={\rm HI,HeI,HeII}}\sum_{i=0}^\infty c_{2i+4}^{(j)} \Lambda^{2i+4} x^{2i+4},
\end{align}
where $\sigma_{\rm T}$ is the Thomson scattering cross section, and we introduced 
\begin{align}
	x &\equiv \frac{h\nu}{k_{\rm B}T_{\rm CMB}} = 0.0176119\left(\frac{\nu}{\rm GHz}\right),\label{def:x}\\
	\Lambda &\equiv \frac{k_{\rm B}T_{\rm CMB}}{a{\rm Ry}} = \frac{1.72663}{a} \times 10^{-5},
\end{align}
with the Rydberg constant Ry, the Boltzmann constant $k_{\rm B}$, the present CMB temperature $T_{\rm CMB}$, the Planck constant $h$, the comoving frequency $\nu$, and the scale factor $a$.
$c_{2i+4}^{(j)}$ are summarized in Tab.~\ref{table2}.
The expansion parameters for HI and HeI are taken from Tab.1 of Ref.~\cite{Alipour:2014dza}.
As discussed in Ref.~\cite{Fi_k_2016}, the Rayleigh scattering cross-section for hydrogen-like atoms of atomic number $Z$ is given as $c^{(\rm HI)}_{2i+4}/Z^{8+4i}$, and 
we consider HeII as a hydrogen-like atom of atomic number $Z=2$.
The differential optical depth is expanded as
\begin{align}
	\dot \tau =\dot \tau^{(0)} +  x^{4} \dot \tau^{(4)}+ x^{6} \dot \tau^{(6)}+ x^{8} \dot \tau^{(8)}+\cdots ,\label{deftau}
\end{align}
where $\dot \tau^{(0)} = -n_{\rm e}\sigma_{\rm T}a$ with the free electron number density $n_{\rm e}$ and  
\begin{align}
	\dot \tau^{(2i+4)} = \dot \tau^{(0)}\sum_{j={\rm HI,HeI,HeII}} \frac{n_{j}}{n_{\rm e}} c_{2i+4}^{(j)}\Lambda^{2i+4}.  \label{eqtau}
\end{align}
$n_{j}$ is the number density of $j$.
We find the number density of ions by modifying a recombination code \texttt{RECFAST}~\cite{Seager:1999bc}.
By default, \texttt{RECFAST} provides hydrogen ionization fraction $x_{\rm H0}$, helium ionization fraction $x_{\rm He0}$, electron ionization fraction $x_{\rm e}$, and helium mass fraction $f_{\rm He}$.
Using those parameters, we get
\begin{align}
	\frac{n_{\rm HI}}{n_{\rm e}} &=x_{\rm e}^{-1}(1-x_{\rm H0}),\label{nHI}\\
	\frac{n_{\rm HeI}}{n_{\rm e}} &=x_{\rm e}^{-1} f_{\rm He}(1-x_{\rm He0})\label{nHeI},
\end{align}
where $n_{\rm H}$ and $n_{\rm He}$ are the total number densities of hydrogen and helium.
Moreover, eliminating the number density of doubly ionized helium $n_{\rm HeIII}$ from 
\begin{align}
		x_{\rm He0}&= \frac{n_{\rm HeII} }{n_{\rm He}}+ \frac{n_{\rm HeIII} }{n_{\rm He}},\\
	x_{\rm e} & = 1- \frac{n_{\rm HI}}{n_{\rm H}} + \frac{n_{\rm HeII}}{n_{\rm H}}+ 2\frac{n_{\rm HeIII}}{n_{\rm H}},
\end{align}
we find
\begin{align}
	\frac{n_{\rm HeII}}{n_{\rm e}} &= x_{\rm e}^{-1}x_{\rm H0} + 2 x_{\rm e}^{-1}f_{\rm He}x_{\rm He0}-1.\label{nheII}
\end{align}
With Eqs.~\eqref{eqtau}, \eqref{nHI} and \eqref{nHeI}, we reproduced the Rayleigh scattering differential optical depth in Refs.~\cite{Lewis:2013yia} and~\cite{Alipour:2014dza} for $\eta>100$Mpc, and we can see a new feature in $\eta<100$Mpc in our Fig~\ref{recfast_out} with Eq.~\eqref{nheII}.
While the number fraction of HeII is smaller than HI, the differential optical depth is comparable to HI because photons are hotter at higher redshift.  
Integrating the differential optical depth at 875GHz, we find $\tau^{(4)}x^4\sim 0.8$, i.e., $80\%$ of photons scatter on $\eta<50$Mpc.
This fact implies that Rayleigh scattering at $\eta<50$Mpc might be important for small scale.
While we mainly use \texttt{RECFAST} in our calculation, we also reproduced Fig.~\ref{recfast_out} using another efficient modern numerical recombination code \texttt{HyRec}~\cite{Ali-Haimoud:2010hou}. We crosschecked the new effect of singly ionized helium Rayleigh scattering.

\section{Rayleigh scattering at first order}
\label{sec3}

We have reviewed background thermodynamics in the previous section.
Now we are ready to compute the evolution of the linear Rayleigh anisotropies.

\subsection{Boltzmann equation}

We will solve the Boltzmann equation for the photon phase-space distribution function~(PDF) in the presence of Thomson and Rayleigh scattering with other cosmological matter contents and the Einstein equation.
It is quite difficult to solve the nonlinear partial differential equations in a general setup, so we perturbatively solve the equation order by order on top of a homogenous and isotropic Friedman background.

The Universe is in local chemical equilibrium in the early epoch, but photon production processes decouple as the Universe expands. The blackbody photosphere is around the redshift $2\times 10^6$ when double Compton scattering becomes inefficient. Thomson scattering and its relativistic correction are dominant afterward until the end of recombination. The photon fluid is defused due to the scattering, and deviation from the Planck distribution is introduced. Therefore, the solution to the photon Boltzmann equation should be written as a local Planck distribution with small spectral distortions.
Indeed, we immediately find the background solution is a homogeneous and isotropic Planck distribution~(See, e.g., \cite{Dodelson:2003ft}) 
\begin{align}
	f_{\rm pl}(x)\equiv \frac{1}{e^x-1},	
\end{align}
with $x$ defined in Eq.~\eqref{def:x}.
%Since $x$ is constant at the background level, the Planck distribution is time-independent.
We perturb the Planck distribution as $f=f_{\rm pl}(xe^{-\Theta})$ or equivalently $T_{\rm CMB}$ as $T_{\rm CMB}e^\Theta$, with the temperature perturbation $\Theta$.
Taylor expanding $f_{\rm pl}(xe^{-\Theta})$ with respect to $\Theta$ up to first order, we find
\begin{align}
    f_{\rm pl}(xe^{-\Theta}) &= f_{\rm pl}(x) + \Theta\mathcal G^{(0)}(x) + \mathcal O(\Theta^2),
\end{align}
where we defined
\begin{align}
	\mathcal G^{(0)}(x) \equiv -x\frac{\partial }{\partial x}f_{\rm pl}(x).
\end{align}
$\Theta$ is frequency independent and is a function of photon momentum direction $\mathbf n$, position $\mathbf x$ and conformal time $\eta$, for Thomson scattering.
$\mathcal G^{(0)}$ is the frequency shape of the first order temperature shift.
Rayleigh scattering shifts the spectrum by $x^{2n+4}$.
Hence, the temperature perturbation for Rayleigh scattering carries additional powers of $x$.
The linearized spectrum including Rayleigh scattering can be parameterized by~\cite{Alipour:2014dza}
\begin{align}
	f = f_{\rm pl} + \mathcal G^{(0)} \Theta^{(0)} + \mathcal G^{(4)} \Theta^{(4)} +\cdots,\label{ansray}
\end{align}
where we include the power of $x$ into %the new frequency dependence as
\begin{align}
	 \mathcal G^{(n)} \equiv  x^{n} \mathcal G^{(0)}.
\end{align}
The dots imply the Rayleigh scattering corrections of $x^{n>4}$.
Rayleigh scattering happens at most 1/10 times during cosmic history for the major frequencies so that the frequency cascade typically happens just once per photon, and the Rayleigh scattered photons are mostly Thomson scattered during the rest of cosmic history.
Therefore, hereafter we truncate series expansions for $x$ at this order unless otherwise stated.
Since $\mathcal G^{(4)}$ perfectly characterizes the frequency dependence at this order, we do not have to consider many samples in the frequency space for calculations~\cite{Alipour:2014dza}.
The Liouville term up to first order in the cosmological perturbation is written as
\begin{align}
    \mathcal L[f] = \left( \Theta^{(0)'}-(\ln x)'\right) \mathcal G^{(0)} + \mathcal G^{(4)} \Theta^{(4)'}.\label{1stlio}
\end{align}
A prime is an ordinary derivative with respect to conformal time.
The time derivatives of functions of $x$ are higher-order perturbations as $(\ln x)'$ is first order in metric perturbations~\cite{Ma:1995ey}.
The linear collision term for Thomson scattering is given as~(See, e.g., \cite{Dodelson:2003ft})
\begin{align}
	\mathcal C[f]&= \dot \tau^{(0)} \left(f - f_0 + x\frac{\partial f}{\partial x}V +\frac{1}{2}f_2 P_2   \right),\label{coleq}
\end{align}
where $V\equiv \mathbf v\cdot \mathbf n$ with the baryon bulk velocity $\mathbf v$, the Legendre polynomial $P_\ell$ of the cosine between $\mathbf n$ and $\mathbf v$, and the multiple coefficient $f_\ell$.  
Following Refs.~\cite{Yu:2001gw,Lewis:2013yia} we obtain the linear Rayleigh collision term by replacing $\dot \tau^{(0)}$ in Eq.~\eqref{coleq} with Eq.~\eqref{deftau}.
Substituting Eqs.~\eqref{ansray} and \eqref{deftau} into \eqref{coleq}, we find
\begin{align}
\mathcal C[f]
=& \dot \tau^{(0)}  \mathcal A^{(0)}\mathcal G^{(0)}+ 
\dot \tau^{(0)}\mathcal A^{(4)}\mathcal G^{(4)}
+
\dot \tau^{(4)}\mathcal A^{(0)}\mathcal G^{(4)},\label{1stcol}
\end{align}
where we introduced
\begin{align}
	\mathcal A^{(n)} &\equiv \Theta^{(n)} - \Theta_0^{(n)} - \delta_{n0} V +\frac{1}{2}\Pi^{(n)} P_2,  \\
	\mathcal A_{\rm P}^{(n)} &\equiv \Theta^{(n)}_{\rm P}  -\frac{1}{2}\Pi^{(n)}(1- P_2).
\end{align}
We included polarization by replacing $\Theta_2$ with $\Pi\equiv \Theta_2+\Theta_{\rm P0}+\Theta_{\rm P2}$.
$\Theta_{\rm P}$ is polarization temperature perturbation, which corresponds to $G_{\gamma}=4\Theta_{\rm P}$ in Ref.~\cite{Ma:1995ey}.

From Eqs.~\eqref{1stcol} and \eqref{1stlio}, the linear order part of the Boltzmann equation $\mathcal L[f] = \mathcal C[f]$ can be obtained as
\begin{align}
	\Theta^{(0)'}-(\ln x)' & = \dot \tau^{(0)}  \mathcal A^{(0)},\label{BT:T}\\
	\Theta^{(4)'}  &= \dot \tau^{(0)}\mathcal A^{(4)}
+
\dot \tau^{(4)}\mathcal A^{(0)}.\label{BT:R}
\end{align}
The first equation is the standard Thomson scattering Boltzmann equation.
The second equation is the first moment of Rayleigh scattering proportional to $\mathcal G^{(4)}$.
The gravitational redshift $(\ln x)'$ does not introduce the Rayleigh anisotropies.
The above derivation straightforwardly applies to the linear Rayleigh polarization.

\subsection{Boltzmann Hierarchy equations}

Linear scalar perturbations associated with the photon PDF are decomposed into
\begin{align}
	\Theta^{(n)} &=(4\pi) \sum_{\ell m}i^\ell \int_{\mathbf k}e^{i\mathbf k\cdot \mathbf x} Y^\star_{\ell m}(\hat k)Y_{\ell m}(\mathbf n)\tilde \Theta_\ell^{(n)} \zeta_{\mathbf k},\label{tenkaitheta}\\
	V &=(4\pi) \sum_{m}i \int_{\mathbf k}e^{i\mathbf k\cdot \mathbf x} Y^\star_{1 m}(\hat k)Y_{1 m}(\mathbf n)\tilde V_1\zeta_{\mathbf k}\label{tenkaiv},
\end{align}
where a tilde implies the transfer function in Fourier space, $\int_{\mathbf k }\equiv \int (2\pi)^{-3} d^3k$ and $\zeta_{\mathbf k}$ is the initial Gaussian stochastic variable that satisfies 
\begin{align}
	\langle \zeta_{\mathbf k}\zeta_{\mathbf k'}\rangle = (2\pi)^3\delta_{\rm D}^{(3)}(\mathbf k+\mathbf k')\frac{2\pi^2}{k^3}\mathcal P_\zeta(k).
\end{align}
$\mathcal P_\zeta$ contains the details of inflationary cosmology.
Expanding the above equations into the Legendre coefficients, we get the Boltzmann hierarchy equations for Rayleigh scattering~\cite{Lewis:2013yia,Alipour:2014dza}
\begin{widetext}
\begin{align}
	\dot{\tilde{\Theta}}^{(4)}_0 &= - k \tilde{\Theta}^{(4)}_{1},\\
	\dot{\tilde{\Theta}}^{(4)}_1 &= \dot\tau^{(0)}\tilde \Theta^{(4)}_1+  \frac{k}{3}\left(\tilde \Theta^{(4)}_{0}- 2 \tilde{\Theta}^{(4)}_{2}\right) + \dot\tau^{(4)}\tilde \Theta^{(0)}_{1g} , \\
	\dot{\tilde{\Theta}}^{(4)}_2 &= \frac{k}{5}\left(2\tilde \Theta^{(4)}_{1}- 3 \tilde{\Theta}^{(4)}_{3}\right) + \dot\tau^{(4)}\left(\tilde \Theta^{(0)}_{2} - \frac{\tilde \Pi^{(0)}}{10}\right)
	+ \dot\tau^{(0)}\left(\tilde \Theta^{(4)}_{2} - \frac{\tilde \Pi^{(4)}}{10}\right),\\
		\dot{\tilde{\Theta}}^{(4)}_{\ell\geq 3} &= \frac{k}{2\ell + 1}\left(\ell \tilde{\Theta}^{(4)}_{\ell-1}- (\ell +1) \tilde{\Theta}^{(4)}_{\ell+1}\right) + \dot\tau^{(4)}\tilde \Theta^{(0)}_\ell + \dot\tau^{(0)}\tilde \Theta^{(4)}_\ell,
\end{align}
where we introduce the photon-baryon velocity difference $\tilde \Theta^{(0)}_{1g}\equiv \tilde{\Theta}^{(0)}_1 - \tilde V_1$.
The polarization hierarchy equations are~\cite{Lewis:2013yia,Alipour:2014dza}
\begin{align}
	\dot{\tilde{\Theta}}^{(4)}_{\rm P,0} &= - k \tilde{\Theta}^{(4)}_{{\rm P,}1} + \dot \tau^{(0)}\left(\tilde \Theta^{(4)}_{{\rm P}0} - \frac{\tilde \Pi^{(4)}}{2}\right) + \dot \tau^{(4)}\left(\tilde\Theta^{(0)}_{{\rm P}0} - \frac{\tilde \Pi^{(0)}}{2}\right),\\
	\dot{\tilde{\Theta}}^{(4)}_{\rm P,1} &= \dot\tau^{(0)}\tilde \Theta^{(4)}_{{\rm P,}1}+  \frac{k}{3}\left(\tilde \Theta^{(4)}_{{\rm P,}0}- 2 \tilde{\Theta}^{(4)}_{{\rm P,}2}\right) + \dot\tau^{(4)}\tilde \Theta^{(0)}_{{\rm P,}1} , \\
	\dot{\tilde{\Theta}}^{(4)}_{\rm P,2} &= \frac{k}{5}\left(2\tilde \Theta^{(4)}_{{\rm P,}1}- 3 \tilde{\Theta}^{(4)}_{{\rm P,}3}\right) + \dot\tau^{(4)}\left(\tilde \Theta^{(0)}_{{\rm P,}2} - \frac{\tilde \Pi^{(0)}}{10}\right)+ \dot\tau^{(0)}\left(\tilde \Theta^{(4)}_{{\rm P,}2} - \frac{\tilde \Pi^{(4)}}{10}\right),\\
		\dot{\tilde{\Theta}}^{(4)}_{{\rm P},\ell\geq 3} &= \frac{k}{2\ell + 1}\left(\ell \tilde{\Theta}^{(4)}_{{\rm P,}\ell-1}- (\ell +1) \tilde{\Theta}^{(4)}_{{\rm P,}\ell+1}\right) + \dot\tau^{(4)}\tilde \Theta^{(0)}_{{\rm P,}\ell} + \dot\tau^{(0)}\tilde \Theta^{(4)}_{{\rm P,}\ell}. 
\end{align}	
\end{widetext}
The hierarchy equations for Thomson scattering are unchanged as Eq.~\eqref{BT:T} does not contain Rayleigh scattering.
The Rayleigh anisotropies and polarization are sourced by $\tilde \Theta_{1g}^{(0)}$, $\tilde \Theta_{\ell>2}^{(0)}$ and $\tilde \Theta_{{\rm P},\ell}^{(0)}$, which are non-vanishing only on sub-horizion scale and are linear gauge invariants.  
The Boltzmann hierarchy equations for Rayleigh scattering are independent of the metric perturbations.
From the above two facts, the Rayleigh anisotropies are gauge independent, and we can use the above hierarchy equations in any gauge.
We truncate the Boltzmann hierarchy with the traditional method in Ref.~\cite{Ma:1995ey}.
In this paper, we numerically evaluate the Boltzmann hierarchy by modifying a publicly available linear Boltzmann solver \texttt{CLASS}~\cite{Blas:2011rf} with cosmological parameters in Ref.~\cite{Aghanim:2018eyx}.

\subsection{Tight-coupling approximation}
\label{sec:TCA}

The Thomson Boltzmann hierarchy is a stiff system as it contains $\dot \tau^{(0)}\tilde \Theta^{(0)}_{1g}$.
Before last scattering, $\dot \tau^{(0)}$ is large, $\tilde \Theta_1$ and $\tilde V_1$ are the almost same size; therefore, we need extra precision for both $\tilde \Theta_{1}$ and $\tilde V_1$ to get sufficiently accurate $\dot \tau^{(0)}\tilde \Theta^{(0)}_{1g}$.
For efficient calculations, we directly find the evolution of $\tilde \Theta_{1g}$ by subtracting the Euler equations for photons and baryons.
The equation is expanded into a series of $1/\dot \tau^{(0)}$ and \texttt{CLASS} truncates the series expansion at second order in $1/\dot \tau^{(0)}$~\cite{Blas:2011rf}.
The multipole of $\ell> 3$ is $\mathcal O(1/\dot\tau^{(0)3})$ and we drop them during the tight coupling approximation.
This simplification significantly reduces the number of equations to solve.

\medskip
The Rayleigh anisotropies are produced by $\tilde \Theta^{(0)}_{1g}$ and higher multipoles $\tilde \Theta^{(0)}_{\ell\geq 2}$.
In contrast to the Thomson hierarchy, the Rayleigh hierarchy is at most $\mathcal O(1/\dot \tau^{(0)2})$.
Up to second order in tight-coupling approximation, we find
\begin{align}
	 \tilde{\Theta}_{1}^{(4)} \approx -\frac{\dot \tau^{(4)}}{\dot \tau^{(0)}} \tilde{\Theta}^{(0)}_{1g},\label{TCA1}
\end{align}
and for $\ell>1$,
\begin{align}
	 \tilde{\Theta}_{\ell}^{(4)} \approx -\frac{\dot \tau^{(4)}}{\dot \tau^{(0)}} \tilde{\Theta}^{(0)}_{\ell}.\label{TCA2}
\end{align}
Similarly, polarization satisfies 
\begin{align}
	 \tilde{\Theta}_{{\rm P},\ell}^{(4)} \approx -\frac{\dot \tau^{(4)}}{\dot \tau^{(0)}} \tilde{\Theta}^{(0)}_{{\rm P},\ell}.\label{TCA3}
\end{align}
For simplicity, we also consider 
\begin{align}
	 \tilde{\Theta}_0^{(4)} = - k\eta \tilde{\Theta}_1^{(4)}.
\end{align}
The expansion scheme for the tight-coupling approximation is nicely reviewed in Ref.~\cite{Blas:2011rf}, and we use the same method.
After the tight coupling approximation, we fully evolve the Rayleigh Boltzmann hierarchy.
%The initial conditions for the full hierarchy equations are obtained as Eqs.~\eqref{TCA1} to \eqref{TCA3}, so 
Adding the new hierarchy equations in the Boltzmann code is straightforward.

\subsection{Back reaction}

Back reaction of Rayleigh scattering appears in the Einstein equation and Euler equation for baryons because those equations contain $x$ integrals of the photon distribution function~\cite{Alipour:2014dza}.
The photon density perturbation is given by
\begin{align}
	\delta_\gamma &= \frac{\int x^3dx  \mathcal G^{(0)} 
	(\tilde \Theta^{(0)}_0+\tilde \Theta^{(4)}_0 x^{4}+\cdots)
	}{\int x^3dx  f_{\rm pl}} 
	\notag \\
	&= 4 \left(\tilde \Theta^{(0)}_0  + \frac{I_{4}}{I_{0}}\tilde \Theta^{(4)}_0 +\cdots \right),
\end{align}
where we have defined
\begin{align}
	I_n \equiv \int dx x^{n+3} \mathcal G^{(0)} =   \zeta (n+4) \Gamma (n+5).
\end{align}
Similarly, we get photon fluid velocity divergence $\theta_\gamma$, photon shear $\sigma_\gamma$, the higher order multipoles $F_{\gamma \ell}$ as
\begin{align}
	\theta_\gamma &= 3k \left(\tilde \Theta^{(0)}_1  + \frac{I_{4}}{I_{0}}\tilde \Theta^{(4)}_1 +\cdots \right), \\
	\sigma_\gamma &= 2  \left(\tilde \Theta^{(0)}_2  + \frac{I_{4}}{I_{0}}\tilde \Theta^{(4)}_2 +\cdots \right), \\
	F_{\gamma,\ell} &= 4\left(\tilde \Theta^{(0)}_\ell  + \frac{I_{4}}{I_{0}}\tilde \Theta^{(4)}_\ell +\cdots \right).
\end{align}
We use the above quantities for the right hand side of the Einstein equation.
The Euler equation for baryons contains the integrated collision term 
\begin{align}
	\dot \tau^{(0)}  (\tilde \Theta_1^{(0)} - \tilde V_1) =  \frac{\int x^3 \left[ \dot \tau^{(0)}  (\tilde \Theta_1^{(0)} - \tilde V_1) \right] \mathcal G^{(0)} dx}{4\int x^3 f_{\rm pl}dx}.
\end{align}
Including Rayleigh scattering, we modify the Thomson optical depth and $\tilde \Theta_1^{(0)}$ and we get
\begin{align}
	\dot\tau^{(0)}\left(\tilde \Theta^{(0)}_1  + \frac{I_{4}}{I_{0}}\tilde \Theta^{(4)}_1 - \tilde V_1\right)
	+\frac{I_4}{I_0} \dot \tau^{(4)} (\tilde \Theta^{(0)}_1 - \tilde V_1).
\end{align}
The back reaction is typically the order of $\mathcal O(I_4/I_0\cdot \dot \tau^{(4)}/\dot\tau^{(0)})$.

\section{Rayleigh scattering at second order}
\label{sec4}

Next, we consider Rayleigh scattering at second order.
To be more precise, we are interested in the spectral distortion of the average CMB spectrum, so we only consider the angular averaged part at second order.
%This is a much simpler situation than the full second-order perturbation theory since we do not have to consider the evolution of the pure second-order perturbations.

\subsection{Collision integral}
\label{sec:col}

The monopole component of the Thomson scattering collision term up to second order is given as~(see e.g., \cite{Ota:2016esq} for a derivation) 
\begin{align}
\mathcal C_0[f]=&-\dot \tau^{(0)} \left[V\left( 3 + x\frac{\partial}{\partial x}\right)f \right]_0
\notag \\
&-\dot \tau^{(0)}  \left[V^2 \left( x^2  \frac{\partial^2}{\partial x^2}
+ 4  x  \frac{\partial}{\partial x}\right)f\right]_0,\label{cft}
\end{align}
where the subscript 0 implies the average with respect to the photon direction $\mathbf n$.
The rest of the second-order collision terms are dropped after the angular average.
The second-order collision term includes the relativistic correction of the bulk motion, which introduces the isotropic spectral distortions at second order~(the Doppler shift).

Since Thomson scattering is an elastic scattering of a photon and an electron, it must rigorously preserve the number of photons for any $f$.
In general, the covariant derivative of the photon number flux $N^{\mu}$ is related to the collision term as~(see, e.g., \cite{Ota:2017jte})
\begin{align}
	\nabla_\mu N^{\mu} = c_1 \int x^2 dx \mathcal C_0[f],
\end{align}
where $c_1$ is a numerical factor of the Jacobian for the perturbed spacetime.
Eq.~\eqref{cft} implies $x^2 \mathcal C_0[f]$ reduces to a total derivative for any $f$ and thus the number flux conservation is always satisfied.

Rayleigh scattering is also elastic and must conserve the number of photons.
However, the straightforward replacement of the differential optical depth in Eq.~\eqref{cft} does not guarantee the number conservation.
The prescription does not work at second order simply because we ignored the incoming photon frequency dependence of Rayleigh scattering~\cite{2018MNRAS.473..457R}.
The non-relativistic limit Rayleigh scattering does not exchange photon energy so that the energies of incoming and outgoing photons are always the same, but this is not the case when including relativistic corrections.

%We may derive the relativistic correction to Rayleigh scattering from quantum electrodynamics.
%The explicit expression is unknown to our knowledge, and the derivation will be complicated.
%This paper fixes the collision terms imposing the number conservation law without the explicit formula.
Derivative operators with respect to $x$ in the collision integral appear when integrating by parts, the Taylor expanded the delta function responsible for the energy conservation of the scattering process.
To be more specific, Eq,~\eqref{cft} is obtained after integrating $\bar x$ out from the following type of integrals~\cite{Ota:2016esq}
\begin{align}
\int d\bar x \dot \tau^{(0)}\left[f(\bar x,\bar{\mathbf n}) - f(x,\mathbf n)\right]
x^n\frac{\partial^n}{\partial \bar x^n}\delta_{\rm D}(\bar x-x)
.\label{cft2}
\end{align}
To properly account for $\bar x$ dependence of the Rayleigh scattering, we include the Rayleigh cross section by replacing the differential optical depth in Eq.~\eqref{cft2} as 
\begin{align}
	\dot \tau^{(0)} \to \dot \tau^{(0)} +  \bar x^{4} \dot \tau^{(4)}+\cdots.\label{deftau2}
\end{align}
With this prescription, substituting the first order solution~\eqref{ansray} and Eq~.\eqref{deftau2} into the second order Thomson collision integral, Eq.~\eqref{cft} yields 
\begin{align}
C_0[f] =& -\dot\tau^{(0)}\left[
V^2  - V\Theta^{(0)}\right]_0 \mathcal Y^{(0)}
&
\notag\\
&+\dot\tau^{(0)}
\left[ V\Theta^{(4)}\right]_0 \mathcal Y^{(4)}
\notag \\
&
 -\dot\tau^{(4)}\left[
V^2 - V\Theta^{(0)}\right]_0 \mathcal Y^{(4)},\label{ct2}
\end{align}
where we introduced
\begin{align}
	\mathcal Y^{(0)} &\equiv \left( -x\frac{\partial }{\partial x}\right)\mathcal G^{(0)} - 3\mathcal G^{(0)},\\
	\mathcal Y^{(n)} &\equiv x^n\mathcal Y^{(0)} - n \mathcal G^{(n)}.
\end{align}
Note that the second-order correction to $f$ turns into third-order correction in Eq.~\eqref{ct2}. 
The first term proportional to $\mathcal Y^{(0)}$ is the source term for the standard $y$ spectral distortion from Thomson scattering~\cite{Chluba:2012gq}.
$\mathcal Y^{(4)}$ is the new spectral distortion.
Both spectral distortions preserve the number of photons as we have
\begin{align}
	\int x^2dx \mathcal Y^{(n)} = 0.
\end{align}

\subsection{Liouville term}
The first-order solution fixes the second-order collision term Eq.~\eqref{ct2}.
Our next step is to find second-order Liouville equation consistent with Eq.~\eqref{ct2}.
For Thomson scattering, second-order ansatz can be written as~\cite{Pitrou:2009bc,Chluba:2012gq,Renaux-Petel:2013zwa,Pitrou:2014ota,Ota:2016esq,Pitrou:2019hqg}
\begin{align}
	f_{\rm pl}\left(xe^{-\Theta^{(0)}}\right) + y^{(0)}\mathcal Y^{(0)}.\label{ydis}
\end{align}
where we ignored $\mu$ spectral distortion since we only consider recombination epoch when $\mu$ is not produced.
The first term is nonlinearly perturbed Planck distribution, and the second term is the $y$ distortion that we cannot include in the temperature.
Eq.~\eqref{ydis} will be generalized to 
\begin{align}
	f_{\rm pl}\left(xe^{-\Theta^{(0)}-x^4\Theta^{(4)}}\right) + y^{(0)}\mathcal Y^{(0)}+ y^{(4)}\mathcal Y^{(4)}.\label{ydis4}
\end{align}
Similar to Thomson scattering, $y^{(0)}$ and $y^{(4)}$ represent the spectral distortion that cannot be included in the frequency dependent temperature perturbation $\Theta^{(0)}+x^4\Theta^{(4)}+\cdots$.
Expanding this equation up to second order, we find
\begin{align}
	f =& f_{\rm pl} + \Theta^{(0)}\mathcal G+ \Theta^{(4)}\mathcal G^{(4)}\notag \\
	 &+\frac32 \Theta^{(0)2}\mathcal G^{(0)}
	 +3 \Theta^{(0)}\Theta^{(4)}\mathcal G^{(4)} 
	 \notag \\
	 &+ \frac12  \Theta^{(0)2}\mathcal Y^{(0)}
	 + \Theta^{(0)}\Theta^{(4)}\mathcal Y^{(4)} \notag \\
	 &+ y^{(0)} \mathcal Y^{(0)}+ y^{(4)} \mathcal Y^{(4)},\label{fexp1}
\end{align}
where we redefine second-order temperature perturbations as $\Theta^{(4)}\to \Theta^{(4)} - 4 y^{(4)}$ for notational simplicity.
Taking the time derivative of Eq.~\eqref{fexp1}, we find
\begin{align}
	&f' = \left(1 +3 \Theta^{(0)}\right)\left(\Theta^{(0)'}- (\ln x)'\right)\mathcal G^{(0)} 
	 \notag \\
	 &
	 +\left[ \left(1 +3 \Theta^{(0)}\right)\Theta^{(4)'}
	 + 3\Theta^{(4)}\left(\Theta^{(0)'}- (\ln x)'\right) 
	 \right]\mathcal G^{(4)}
	  \notag \\
	 &
	 	+[ y^{(0)'} 
	 	+  \Theta^{(0)} \left(\Theta^{(0)'}- (\ln x)'\right)
	 	]
	 	\mathcal Y^{(0)} 
	  \notag \\
	 &+[y^{(4)'}+ \Theta^{(0)}\Theta^{(4)'}+ \Theta^{(4)}\left(\Theta^{(0)'}- (\ln x)'\right) ]\mathcal Y^{(4)}, \label{li2}
\end{align}
where we used the chain rule:
\begin{align}
	\mathcal G^{(n)'} &=  -(\ln x)'\left(\mathcal Y^{(n)} + 3\mathcal G^{(n)}
	\right).
\end{align}
We do not have to consider $\mathcal Y^{(n)'}$ since this term turns into a third-order correction that we ignore.

\begin{figure*}
  \includegraphics[width=0.45\linewidth]{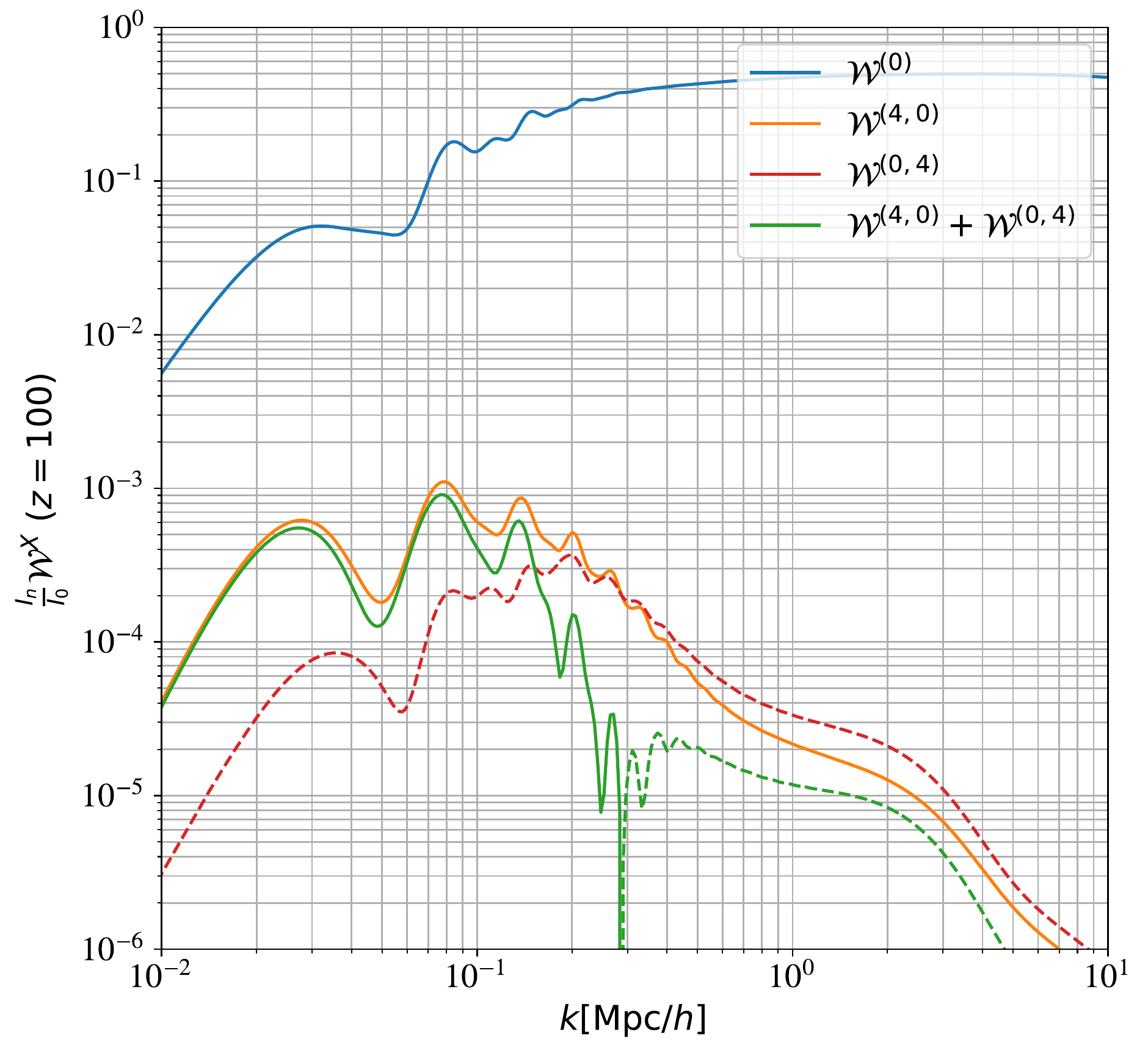}
      \includegraphics[width=0.45\linewidth]{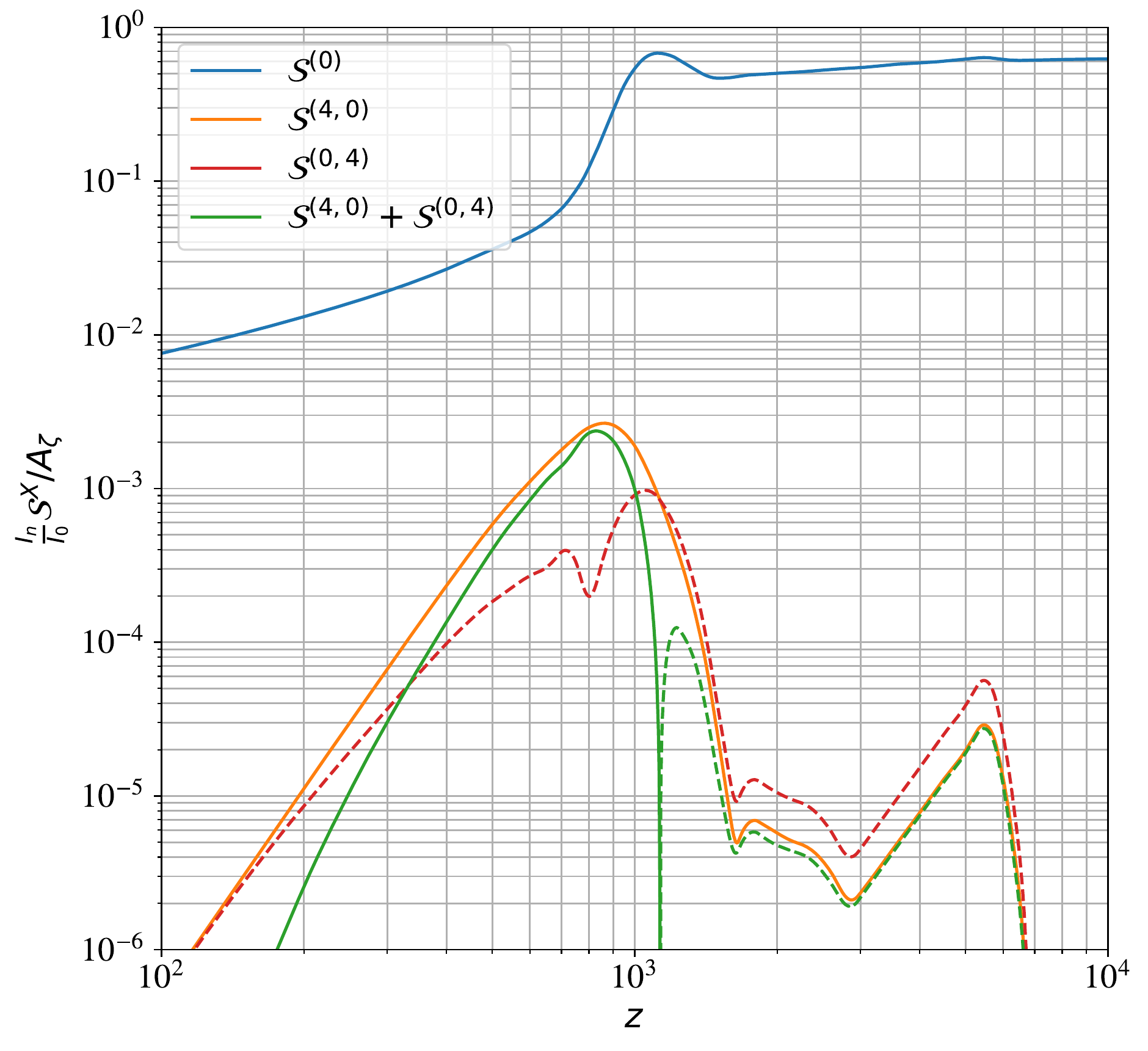}
      
       \caption{\textit{Left}: the Fourier space window function $I_n/I_0\cdot \mathcal W^{(X)}$ for the spectral distortions at $z=100$. The dashed curves imply the negative sign. $\mathcal W^{(0,4)} \approx -2 \mathcal W^{(4,0)}$ so that the window function is negative during the tight coupling regime, but $\mathcal W^{(4,0)}>0$ dominates after hydrogen recombination.
       \textit{Right}: the spectral distortion source $I_n/I_0\cdot \mathcal S^{(X)}$ for the primordial power spectrum of spectral index $n_s=0.96$. Similarly, $\mathcal S^{(0,4)} \approx -2 \mathcal S^{(4,0)}$ is established for high redshifts, but $\mathcal S^{(4,0)}>0$ dominates after hydrogen recombination.
       }
    \label{window}
\end{figure*}

\subsection{Boltzmann equation}
Equating Eqs.~\eqref{ct2} and~\eqref{li2}, with the linear collision terms \eqref{BT:T} and \eqref{BT:R}, we find

	\begin{align}
	\Theta^{(0)'}_0=&-3\dot \tau^{(0)} [\Theta^{(0)}\mathcal A^{(0)}]_0, \\
	\Theta^{(4)'}_0=&-3\dot \tau^{(0)} [\Theta^{(0)} \mathcal A^{(4)}]_0
	 -3\dot \tau^{(4)}[\Theta^{(0)} \mathcal A^{(0)}]_0
	\notag \\
	&-3\dot \tau^{(0)}[\Theta^{(4)} \mathcal A^{(0)}]_0, \\
	y^{(0)'}_0=&-\dot\tau^{(0)}[\Theta^{(0)}\mathcal A^{(0)}]_0-\dot\tau^{(0)}\left[
V^2  - V\Theta^{(0)}\right]_0, \label{y0dis}\\
	y^{(4)'}_0=&-\dot\tau^{(0)}[\Theta^{(4)}\mathcal A^{(0)}]_0-\dot\tau^{(0)}[\Theta^{(0)}\mathcal A^{(4)}]_0
		\notag \\
		&+\dot\tau^{(0)}
\left[ V\Theta^{(4)}\right]_0 
	-\dot\tau^{(4)}[\Theta^{(0)}\mathcal A^{(0)}]_0
	\notag \\
&
-\dot\tau^{(4)}\left[
V^2  - V\Theta^{(0)}\right]_0.\label{y4dis}
\end{align}

The source terms for $y^{(0)}_0$ and $y^{(4)}_0$ are given as products of the linear scalar perturbations.
The ensemble and angular average of a product of linear perturbations is evaluated as 
\begin{align}
	\langle[ AB]_0\rangle  =&\int d\ln k \mathcal P_\zeta \sum_{\ell}  (2\ell+1) \tilde A_{\ell}\tilde B_{\ell}.  \label{lltenkai}
\end{align}
Using Eqs.~\eqref{tenkaitheta}, \eqref{tenkaiv},~\eqref{y0dis},~\eqref{y4dis}, and \eqref{lltenkai},
we find
\begin{align}
	\langle y^{(0)}_0\rangle' &= \int d\ln k \mathcal P_\zeta \mathcal K^{(0)},\label{y0source}\\
	\langle y^{(4)}_0\rangle ' &= \int d\ln k \mathcal P_\zeta \left[\mathcal K^{(0,4)}+\mathcal K^{(4,0)}\right],\label{ysource}
\end{align}
where we defined
\begin{align}
	-\frac{\mathcal K^{(0)}}{\dot\tau^{(0)}} =&3\tilde \Theta^{(0)2}_{1g}  -\frac{\tilde \Theta^{(0)}_2 \tilde \Pi^{(0)}}{2}
	+ \sum_{\ell=2}^\infty(2\ell+1) \tilde \Theta^{(0)2}_\ell,\\
	 -\frac{\mathcal K^{(0,4)}}{\dot\tau^{(0)}}  =& 
	6\tilde \Theta^{(4)}_1 \tilde \Theta^{(0)}_{1g}
	- \frac{\tilde \Theta^{(4)}_2 \tilde \Pi^{(0)}}{2}
	- \frac{\tilde \Theta^{(0)}_2 \tilde \Pi^{(4)}}{2}
	\notag \\
	&
	+2\sum_{\ell=2}^\infty(2\ell+1)\tilde \Theta^{(4)}_\ell \tilde \Theta^{(0)}_\ell \label{k04}
	\\
	\mathcal K^{(4,0)} =& \frac{\dot\tau^{(4)}}{\dot\tau^{(0)}}\mathcal K^{(0)}.\label{k40}
\end{align}
The gauge invariance of spectral distortions is critical as the spectral distortions are physical observables. 
The kernel function for the traditional $y$ spectral distortion is composed of $\tilde \Theta_{1g}^{(0)}$ and $\tilde \Theta^{(0)}_{\ell\geq 2}$.
Those variables are gauge invariant at linear order, so $y$ distortion is also gauge invariant~\cite{Chluba:2012gq}.
The new spectral distortion is sourced by $\tilde \Theta_{1g}^{(0)}$, $\tilde \Theta^{(0)}_{\ell\geq 2}$, and $\tilde \Theta^{(4)}_{\ell\geq 1}$.
While $\tilde \Theta^{(0)}_{1}$ is gauge dependent, $\tilde \Theta^{(4)}_{1}$ is not since it is sourced by $\tilde \Theta_{1g}^{(0)}$.
Therefore, the new spectral distortion is also gauge invariant.

Energy release at high redshift is thermalized due to the energy transfer by high energy electrons.
$y$ and our new spectral distortions are realized when the thermalization effect is weak.
Roughly speaking, the spectral dependence of the energy release after $z=6\times 10^4$ is not changed anymore, and Ref.~\cite{Chluba:2013vsa} found a precise energy branching ratio
\begin{align}
	\mathcal J_y(z) = \frac{1}{1+\left(\frac{1+z}{6.0\times 10^4}\right)^{2.58}},
\end{align}
and, $\mathcal J_y \mathcal K^{(X)}$ contributes to the new spectral distortion.
Integrating Eq.~\eqref{ysource} with respect to time, we finally get
\begin{align}
	\langle y^{(4)}_0\rangle  = \int_{k_{\rm min}}^{k_{\rm max}} d\ln k \mathcal P_\zeta \left( \mathcal W^{(0,4)}
	+
	\mathcal W^{(4,0)}\right),
\end{align}
where we introduced the window function
\begin{align}
	\mathcal W^{(X)}(k,\eta) & \equiv \int^\eta_{0} d\bar \eta \mathcal J_y \mathcal K^{(X)}.
\end{align}
The Fourier space window functions~$\mathcal W^{(X)}$ are presented in the left panel of Fig.~\ref{window}.
In the figure, we normalized the window function by the energy fraction of the new spectral distortion
\begin{align}
	\frac{\int x^3 dx \mathcal Y^{(n)}}{\int x^3 dx \mathcal Y^{(0)}} = \frac{\int x^3 dx \mathcal G^{(n)}}{\int x^3 dx \mathcal G^{(0)}} = \frac{I_n}{I_0}.
\end{align}

\medskip
Assuming standard power law initial power spectrum
\begin{align}
	\mathcal P_\zeta = A_\zeta \left(\frac{k}{k_0}\right)^{n_s-1},
\end{align}
with $k_0{\rm Mpc}/h=0.05$ and $n_s=0.96$, we integrate the Fourier momentum in Eq.~\eqref{ysource} and find
\begin{align}
	\langle y^{(4)}_0\rangle  & = \int_z^{\infty} d \ln(1+z) \left( \mathcal S^{(0,4)} + \mathcal S^{(4,0)}\right),
\end{align}
where
\begin{align}
		\mathcal S^{(X)}(z) &\equiv \int_{k_{\rm min}}^{k_{\rm max}} d\ln k \mathcal P_\zeta  \mathcal J_y\frac{\mathcal K^{(X)}}{H}. 
\end{align}
The source functions $\mathcal S^{(X)}$ in units of $A_\zeta$ as a function of the redshift are shown in the right panel of Fig.~\ref{window}.

\subsection{Acoustic dissipation}

Rayleigh scattering is an additional photon scattering that makes photon-baryon coupling stronger.
Hence, Rayleigh scattering decreases the multipole anisotropy during the tight coupling regime.
As a result, the energy released from acoustic oscillation is reduced.
Thus, the new spectral distortion is a negative contribution to the $y$ distortion during the tight coupling regime.
%At $k{\rm Mpc}/h\gtrsim 1$, the tight coupling approximation is valid and $\mathcal W^{(0,4)} \approx -2 \mathcal W^{(4,0)}$ is satisfied.
%Also $\mathcal S^{(0,4)} \approx -2 \mathcal S^{(4,0)}$ for $z\gtrsim 1500$.
%During the tight-coupling regime, 
Combining Eqs.~\eqref{TCA1}, \eqref{TCA2}, \eqref{TCA3} and \eqref{k04}, we find 
\begin{align}
	\langle y^{(4)}_0\rangle ' \sim -\frac{\dot \tau^{(4)}}{\dot \tau^{(0)}} \langle y^{(0)}_0\rangle',
\end{align}
which is consistent with both panels in Fig.~\ref{window}.
Including Rayleigh scattering effectively implies increasing baryons for Thomson scattering, and we also confirmed $\partial y^{(0)}/\partial \Omega_b<0$ during recombination.
%$\mathcal S^{(0,4)}$ and $\mathcal W^{(0,4)}$ represent Thomson scattering of the Rayleigh anisotropies, and $\mathcal S^{(4,0)}$ and $\mathcal W^{(4,0)}$ imply Rayleigh scattering of the Thomson anisotropies.
%After hydrogen recombination, Thomson scattering becomes inefficient 
%so that the Rayleigh anisotropies are not Thomson scattered anymore. As a result, the cross term $\mathcal W^{(0,4)}$ and $\mathcal S^{(0,4)}$ are reduced.
%On the other hand, the Thomson anisotropies can Rayleigh scatter after hydrogen recombination; therefore, $\mathcal W^{(4,0)}$ and $\mathcal S^{(4,0)}$ become important after hydrogen recombination.

After hydrogen recombination, Thomson scattering is less efficient.
Then, Rayleigh scattering adds extra diffusion, which causes a positive spectral distortion. 
Energy contribution from the final recombination stage is dominant because hydrogen Rayleigh scattering is most significant.
For $n_s=0.96$ and $A_\zeta=2.196\times 10^{-9}$, the main contribution comes from $k{\rm Mpc}/h\sim 0.1$, and the total new spectral distortion has the same sign as $y$.
We found a new spectral distortion from acoustic dissipation during recombination is $y^{(4)}_{\rm ac.}\mathcal Y^{(4)}= 4\times 10^{-3}$Jy/str around $\nu = 300$GHz and 600GHz in Fig.~\ref{spectrum}.
We also confirmed that the $y$ distortion decreases by $0.1\%$, including the backreaction effect. 
Those corrections are one order of magnitude smaller than the sensitivity range of ESA voyage 2050~\cite{Chluba:2019kpb,Chluba:2019nxa}.

\subsection{Reionization}

\begin{figure}
\centering
  \includegraphics[width=\linewidth]{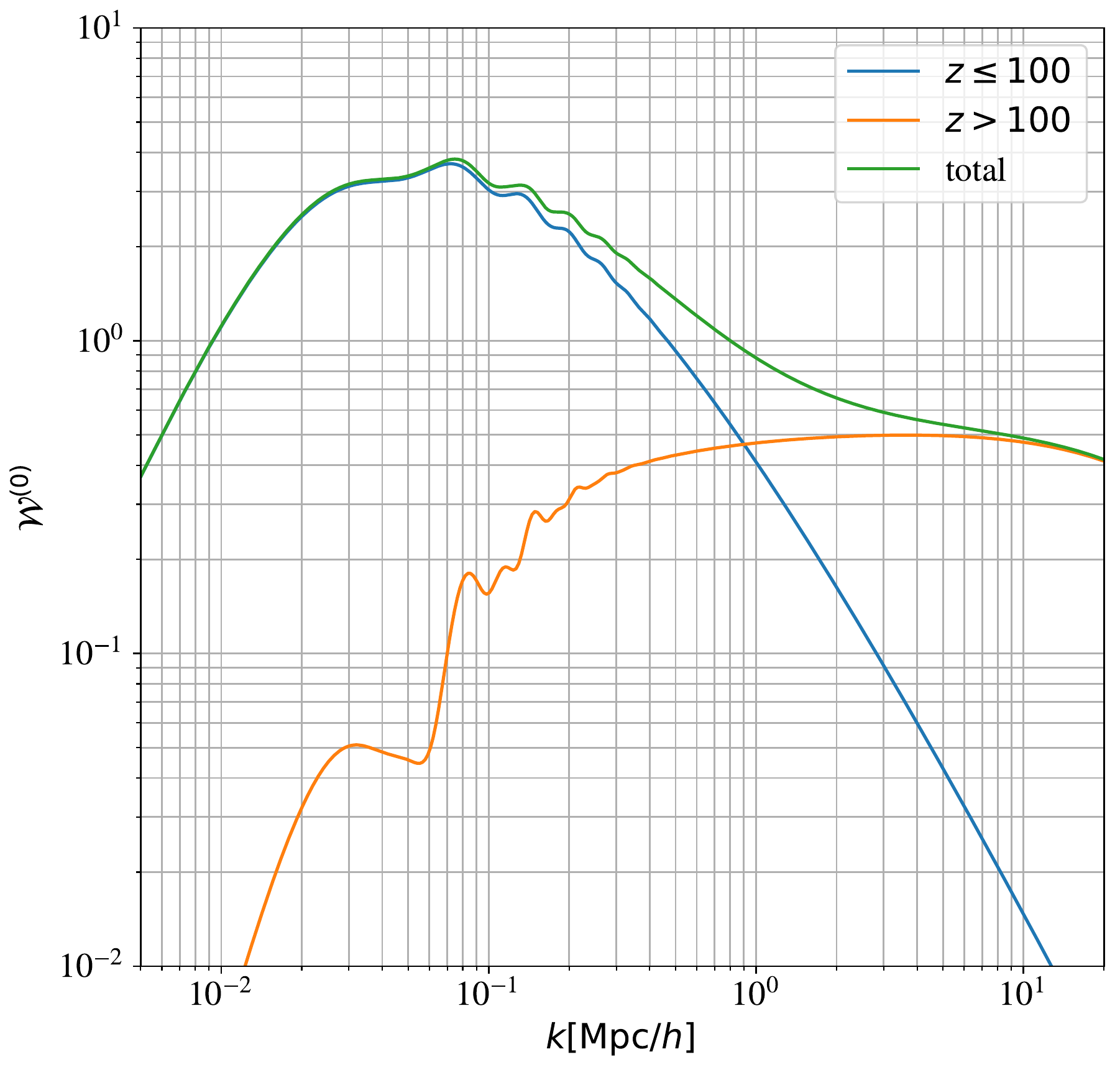}
  \caption{The $y$ distortion window functions for $z\leq100$ and $z>100$.
  The window function for $z\leq 100$ is amplified because the baryon velocity is growing with the dark matter sector. Such a reionization enhancement does not happen for Rayleigh scattering.   
  }
	\label{reio_window}
\end{figure}

After hydrogen recombination, both Thomson scattering and Rayleigh scattering are inefficient.
Photons free-stream while baryons evolve like dark matter after recombination.
Reionization introduces additional $y$ spectral distortion since the source term is enhanced drastically as $V_1\gg\Theta_1^{(0)}$ at a late time.
$y$ distortion from the reionization is five times bigger than recombination.
The Fourier space window functions for $z\leq 100$ and $z>100$ are shown in Fig.~\ref{reio_window}.
Including reionization contribution $y^{(0)}_{\rm reio} =2.4\times10^{-8} $, we find the total spectral distortion $y^{(0)}=2.9\times 10^{-8}$ at present.
Rayleigh scattering will not happen at a late time since the photon energy is too small; therefore, $\mathcal S^{(4,0)}$ is negligible for reionization.
The Rayleigh anisotropies produced during recombination can also be Thomson scattered after reionization.
However, the Rayleigh dipole $\Theta_{1}^{(4)}$ is not enhanced, unlike $\Theta_{1g}^{(0)}$ since it does not have the baryon velocity.
Hence, $\mathcal S^{(0,4)}$ is also negligible.
We evaluated the spectral distortion from Rayleigh scattering after reionization, and we found that it is 0.1\% of the recombination one.
Thus, Rayleigh scattering will not generate secondary spectral distortions during reionization.

\subsection{Compton cooling}
So far, we have ignored the relativistic correction $T_{\rm e}/m_{\rm e}$ to the Boltzmann equation.
This term leads to the relativistic correction to the isotropic part of the photon distribution function independently from second-order perturbations.
The relativistic correction to the Thomson scattering collision integral is~(see e.g., \cite{Ota:2016esq} for a derivation)
\begin{align}
&-\dot \tau^{(0)}  \frac{T_{\rm e}}{m_{\rm e}} \left( x^2  \frac{\partial^2}{\partial x^2}
+ 4  x  \frac{\partial}{\partial x}\right)f
\notag \\
&
-\dot \tau^{(0)} \frac{T_{\rm CMB}(1+z)}{m_{\rm e}}  \left[x^2\frac{\partial }{\partial x} +4x\right]f(1+f),
\label{cfte}
\end{align}
and for $f=f_{\rm pl}$, Eq.~\eqref{cfte} turns into 
\begin{align}
	 -\dot \tau^{(0)} \left[ \frac{T_{\rm e}}{m_{\rm e}} -\frac{T_{\rm CMB}(1+z)}{m_{\rm e}} \right] \mathcal Y^{(0)}.
\end{align}
The $y$ distortion from this term is called Compton cooling as the electron temperature is always slightly lower than the photon's, and photons are cooled via Compton scattering~\cite{Chluba:2011hw}.
Including reionization, we found $y_{\rm e}^{(0)}=2.74\times 10^{-10}$, which is subdominant to acoustic dissipation.
With the same prescription as Sec.~\ref{sec:col}, we found $\mathcal Y^{(0)}$ is replaced with $\mathcal Y^{(4)}$ for Rayleigh scattering and find the following additional contribution
\begin{align}
	 -\dot \tau^{(4)} \left[ \frac{T_{\rm e}}{m_{\rm e}} -\frac{T_{\rm CMB}(1+z)}{m_{\rm e}} \right] \mathcal Y^{(4)}.
\end{align}
Since $\tau^{(4)}$ is suppressed during reionization, the Compton cooling via Rayleigh scattering only happens during recombination.
The cooling effect for Rayleigh scattering is estimated as $y^{(4)}_{\rm e}\mathcal Y^{(4)}\sim 2.5\times 10^{-3}$Jy/str.
As Rayleigh scattering introduces an additional cooling effect, the spectral distortion has the same sign as the standard cooling effect.
$y^{(0)}_{\rm ac}$ is sourced during recombination, but $y^{(0)}_{\rm e}$ is produced only at the end of recombination. As a result, we find $y^{(0)}_{\rm e}\ll y^{(0)}_{\rm ac}$.
On the other hand, both $y^{(4)}_{\rm ac}$ and $y^{(4)}_{\rm e}$ are sourced at the end of recombination; those signals are comparable.
The total $y^{(4)}$ increases by about 60\% after including the cooling effect.

\subsection{Sunyaev-Zel'dovich effect}
$y$ spectral distortion from recombination is useful observable for tests of primordial perturbations at $0.1 \leq k{\rm Mpc}/h \leq 100$~\cite{Chluba:2012gq,Chluba:2012we}.
However, $\mathcal Y^{(0)}$-type spectral dependence is dominated by the thermal Sunyaev-Zel'dovich effect, i.e., the Compton scattering between the CMB photons and hot electron gas in the late Universe.
The typical size of $y^{(0)}_{\rm tSZ}$ is $10^{-6}$~\cite{Refregier:2000xz}, which is 100 times bigger than the primordial signals even after including reionization enhancement.
The tSZ effect is indistinguishable from reionization or recombination signals.
The degeneration is a crucial disadvantage of $y$ spectral distortion for primordial perturbation search.
While the new spectral distortion from Rayleigh scattering is weak, Rayleigh scattering must not secondarily produce the signal after recombination because the photons must be sufficiently hot.
For clusters at $z<10$, the suppression factor is below $10^{-8}$.
Therefore, the new spectral distortions are straightforwardly related to cosmology during recombination.

\section{Constraints on cosmology}
\label{sec5}

\begin{figure}
\centering
  \includegraphics[width=\linewidth]{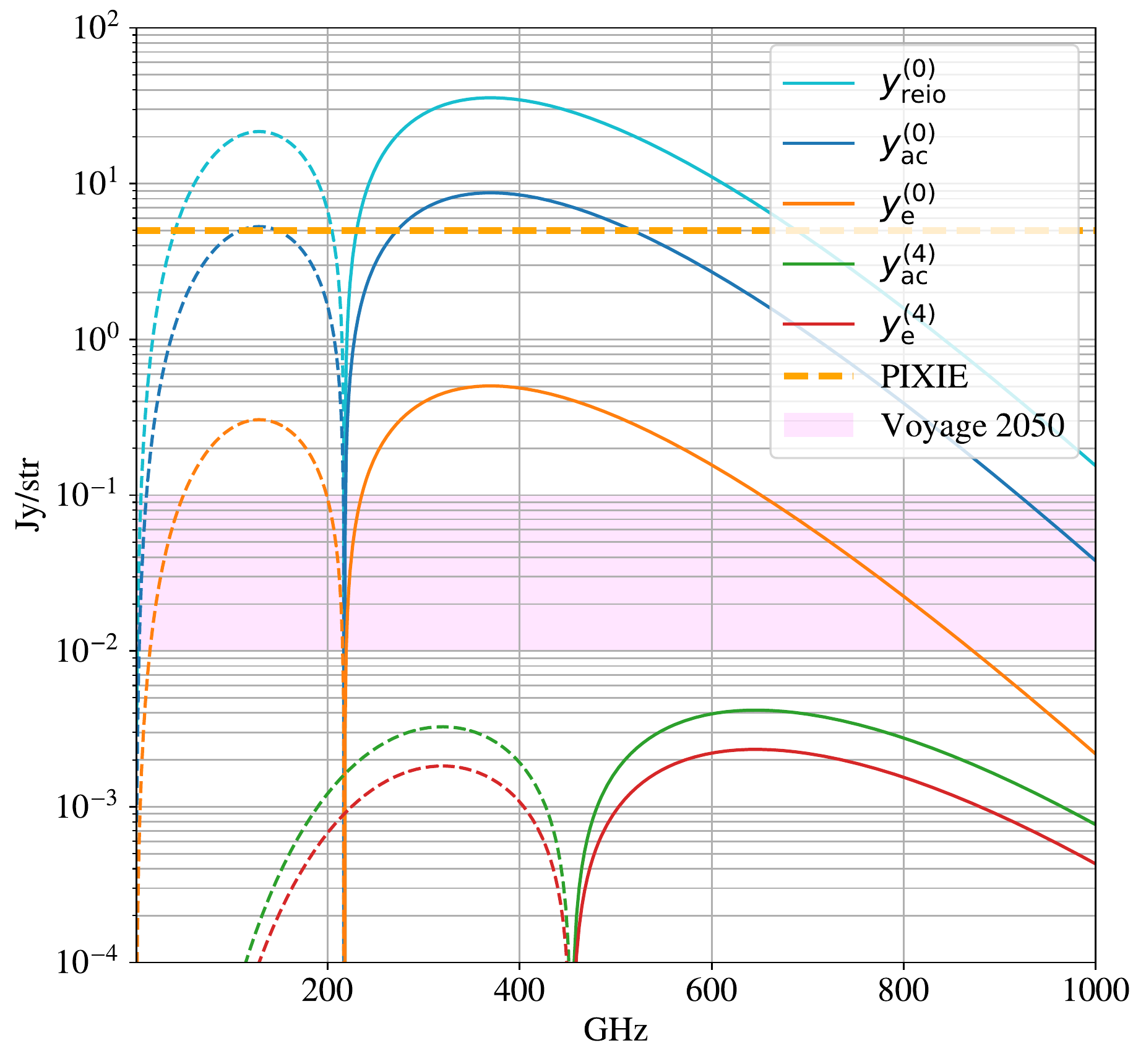}
  \caption{
  The $y$ spectral distortion and the new spectral distortion from Rayleigh scattering with future experimental target sensitivities.
 The spectral distortions are evaluated for cosmological parameters consistent with the recent CMB anisotropy measurements. 
  }
	\label{spectrum}
\end{figure}

We calculated the new spectral distortions for the standard power-law initial spectrum and found the total spectral distortion, including the cooling effect, is $6.5\times 10^{-3}$Jy/str.
The size is about 0.7\% of the $y$ spectral distortion from recombination and one order of magnitude smaller than the target sensitivity range of voyage 2050~\cite{Chluba:2019kpb,Chluba:2019nxa}.
Observing the new spectral distortions for the standard cosmological scenario with the almost scale invariant curvature perturbatins is quite difficult.
However, there are a few advantages of the new spectral distortions.
The $y$ spectral distortion degenerates with the tSZ effect, which is $10^3$ times bigger than the $y$ distortion from recombination.
Therefore, we cannot use $y^{(0)}$ for cosmological constraints about recombination even if the sensitivity goes beyond $10^{-3}$Jy/str.
The new spectral distortion is free from the issue.
We can extract the cosmological information from ideal measurements of $y^{(4)}$ and crosscheck cosmology determined by other surveys such as Planck~\cite{Aghanim:2018eyx}.

There are two peaks in the right panel of Fig.~\ref{window}.
The first negative peak around $z\sim 6000$ is due to Rayleigh scattering of singly ionized helium during the tight-coupling regime.
The second positive peak is from neutral hydrogen Rayleigh scattering.
The late time peak dominates the total distortion for the almost scale-invariant power spectrum because hydrogen Rayleigh scattering is more efficient than singly ionized helium. 

The scale dependence of the sign could be potentially useful to distinguish several nonstandard scenarios.
For example, let us consider a blue tilted initial spectrum, i.e., $n_s>1$.
As we have $\log_{10}[W^{(0,4)}(k{\rm Mpc/h}=1.)/\mathcal W^{(4,0)}(k{\rm Mpc/h}=0.1)]\approx 3.4$, the negative part dominates for the blue spectrum with $n_s\gtrsim 4.4$.
However, such a highly blue spectrum also introduces detectable spectral distortions like $\mu$ spectral distortions~\cite{Cabass:2016giw}.
Several inflationary models predict peaks in the curvature power spectrum that can be approximated by the lognormal spectrum
\begin{align}
	\mathcal P_\zeta = \frac{A_\zeta}{\sqrt{2\pi}\Delta} e^{-\frac{\left(\ln k - \ln k_0\right)^2}{2\Delta^2}},
\end{align}
which is often considered for primordial black hole formation and associated induced gravitational waves~\cite{Pi:2020otn}.
Consider the peak located at $k_0{\rm Mpc}/h=2.$, with $A_\zeta = 10^{-4}$ and $\Delta=10^{-3}$.
We cannot exclude such a lognormal spectrum from the current temperature anisotropy measurements since the CMB anisotropies at $k{\rm Mpc}/h>0.1$ is exponentially suppressed due to Silk damping.  
In this setup, we find $y^{(0)}_{\rm ac}=1.\times 10^{-6}$, which is comparable to the tSZ effect, so we cannot distinguish the tSZ effect from $y_{\rm ac.}^{(0)}$.
For the same parameters, we find the new spectral distortion is $3.\times 10^{-2}$Jy/str with the opposite sign, which is in the target range of voyage 2050.

\section{Conclusions}
\label{sec6}

Towards future precise CMB intensity spectral measurements~\cite{Kogut:2011xw,Kogut:2019vqh,PRISM:2013fvg,Chluba:2019kpb,Chluba:2019nxa}, we considered a new spectral distortion from Rayleigh scattering during recombination.
There are a few works on the Rayleigh anisotropies~\cite{Yu:2001gw,Lewis:2013yia,Alipour:2014dza,Beringue:2020wxw,Coulton:2020oxw}, but all of them are about the linear anisotropies, and thus distortions to the angular averaged spectrum have not been considered.
In this work, we expanded the photon Boltzmann equation to second-order cosmological perturbations in the presence of Thomson scattering and Rayleigh scattering, and we show that the second-order effect introduces a new monopole spectral distortion.
The generation mechanism is similar to the traditional $y$ spectral distortion from the relativistic corrections to the Thomson scattering.
We show that Rayleigh scattering also introduces the frequency-dependent acoustic dissipation and Compton cooling effect. 
The spectral shape is different from the other traditional spectral distortions.
Spectral distortions from atomic processes during recombination were studied at the background level in the literature, but those works do not include Rayleigh scattering since diffusion introduced by second order perturbations is mandatory for the present consideration~\cite{Wong:2005yr,Sunyaev:2009qn,Chluba:2015gta,Chluba:2008aw}.

Fig.~\ref{spectrum} shows several $y$ and new spectral distortions obtained in this work.
We found the new spectral distortion, including both acoustic dissipation and the cooling effect, is $6.5\times 10^{-3}$Jy/str for the cosmological parameters determined by Ref.~\cite{Aghanim:2018eyx}, which is one order of magnitude smaller than the envisioned target sensitivity range of voyage 2050.
Thus, the spectral distortion from the second-order Rayleigh scattering is tiny, and it will not be easy to detect the signal soon.

Even if the new spectral distortion is tiny, it can be useful.
The scale invariant primordial density perturbations at $0.01\leq k {\rm Mpc}\leq 100$ yields $y \sim 10^{-9}$, which is the detectable size in the next generation of spectral measuremets.
However, the Sunyaev-Zel'dovich effect is known as the same spectral dependence as the $y$ distortion; Compton scattering of the CMB photons by high energy electron gas in the late time clusters is three orders of magnitude bigger than the primordial $y$ distortion.
In principle, the sky averages degenerate.
The new spectral distortion is not produced in the late Universe because CMB photons are too cold for Rayleigh scattering after recombination.
In the longer term, the new spectral distortion can be distinguishable from other signals and can be more useful than the traditional $y$ distortion for recombination studies.
We also considered several scale-dependent primordial power spectra for the new spectral distortion, and we showed that there is a unique window for constraining the large initial perturbations at $k{\rm Mpc}/h\sim 1$.

Rayleigh scattering adds new electrons interacting with photons; therefore, the photon-electron coupling becomes stronger.
As a result, the multipole anisotropies are suppressed, and then the acoustic dissipation is reduced during the tight coupling regime.
Thus, additional interacting species suppress the acoustic source in the early Universe.
Such a response will be universal; therefore, spectral distortions will be useful for constraining unknown interacting dark sectors.

We considered a rigorous framework of second-order Boltzmann equations for isotropic spectral distortions. 
Full second-order CMB, including Rayleigh scattering, will be more complicated but important to subtract the intrinsic nonlinearity for constraining the primordial non-Gaussianity in the Rayleigh anisotropies.
The extension to the full second-order CMB will be discussed elsewhere.

\begin{acknowledgments}

The author would like to thank William R. Coulton for his useful discussions.

\end{acknowledgments}

\appendix

\bibliography{bib}{}
\bibliographystyle{unsrturl}

\end{document}